# Neutrino Oscillations in Structured Matter

Paul M. Fishbane

*Physics Dept. and Institute for Nuclear and Particle Physics,*
*Univ. of Virginia, Charlottesville, VA 22903*

## Abstract

A layered material structure in a monochromatic neutrino beam produces interference effects that could be used for the measurement of features of the neutrino mass matrix. The phenomenon would be most useful at high energies.

The MSW effect [1, 2] describes how electron neutrinos in matter propagate differently from other neutrinos, and from electron neutrinos in vacuum. This effect is an element in the interpretation of recent experiments [3-6] that have explored the neutrino mass spectrum. The phenomenon also describes the effects of the presence of boundaries between different media on neutrino propagation and oscillation. As recent work [7-9] has shown, the boundaries introduce the possibility of interference between different amplitudes for neutrino propagation. Indeed, the presence of boundaries within the earth has implications [7, 8] for the interpretation of the data of reference 3.

Here we point out that an arrangement of layers of materials, containing many boundaries, can provide another angle on the interference between the propagation modes. In addition to describing the basic mechanism for a two-family neutrino structure, we briefly address the question of where this effect could be most profitably employed.

## I. Review of MSW and Boundary Interference effects

Neutrino oscillation occurs because the electroweak eigenstates are not the mass eigenstates. Let us consider two families of neutrinos, with electroweak labels $e$ and $\mu$ (generically Greek letters) and mass eigenstate labels 1 and 2, with 1 labeling the lightest neutrino. The matrix $U$ (the leptonic CKM matrix) connects these states according to

$$\left| \nu_\alpha \right\rangle = \sum_{i=1,2} U_{\alpha i} \left| \nu_i \right\rangle, \tag{1.1}$$

where $U$ takes the generic form

$$U = \begin{bmatrix} \cos\theta & \sin\theta \\ -\sin\theta & \cos\theta \end{bmatrix}. \tag{1.2}$$

with $\theta$ such that $\cos(2\theta)$ is positive.

Starting with a pure beam of, say, $\nu_\beta$, with definite momentum (assumed throughout), the time evolution is governed by the mass eigenstates and gives after the neutrino beam has traveled a distance $x \cong t$ a beam containing a mixture of each type of neutrino in the usual fashion. In particular the probability for the conversion $\beta \to \alpha$ is

$$P_{\beta\alpha}(t) = \sin^2 2\theta \sin^2 \varphi, \tag{1.3}$$

where the angle $\varphi$ is determined by the energy difference,

$$\varphi = \frac{1}{2} \left( E_2 - E_1 \right) x \cong \frac{\Delta m^2}{4E} x, \tag{1.4}$$

with $\Delta m^2 \equiv m_2{}^2 - m_1{}^2$. The approximation refers to the limit in which the mass difference $\Delta m$ is much less than the momentum of the beam, a limit that will be of interest to us. The oscillation length $\ell(E)$ is the distance that corresponds to a change in $\varphi$ by $\pi$; that is $\ell(E) = 4\pi E/\Delta m^2$. Finally note that the probability for nonconversion, i.e. that $\beta \to \beta$, is

$$P_{\beta\beta} = \cos^2 \varphi + \cos^2 2\theta \sin^2 \varphi = 1 - P_{\beta\alpha}. \tag{1.5}$$

***Propagation in matter***. In the presence of matter, each neutrino specie may have a different potential. (In normal matter it is only the electron neutrinos for which there is a potential associated with rescattering from electrons in the material.) In particular, suppose that the potential difference of the two neutrino species is $V_k$, where $k$ labels the material. Then the effect of the matter on the propagation parameters is described by

$$\Delta m^2 \rightarrow \Delta m_k^2 = \Delta m^2 \sqrt{\left(\cos 2\theta - \xi_k\right)^2 + \sin^2 2\theta}$$

$$\varphi \rightarrow \varphi_k = \frac{\Delta m_k^2}{4E} x_k \tag{1.6}$$

$$\theta \rightarrow \theta_k \ni \sin^2 2\theta_k = \frac{\sin^2 2\theta}{\left(\cos 2\theta - \xi_k\right)^2 + \sin^2 2\theta}$$

where

$$\xi_k \equiv \frac{2EV_k}{\Delta m^2} \tag{1.7}$$

The value of $E$ for which $\xi_i = \cos 2\theta$ is referred to as the MSW resonance; for that value the angle $2\theta_i$ goes through $\pi/2$. for general values of $\theta$, total conversion in a single thickness of any medium, including vacuum, is not possible, but it can occur for an appropriate thickness of medium at MSW resonance.

The expression for $\theta_i$ above is equivalent to the form given in Ref. [2], namely

$$\tan 2\theta_i = \frac{\tan 2\theta}{1 - \dfrac{4\pi E}{\ell_0 \Delta m^2} \sec 2\theta} \tag{1.8}$$

with the replacement $\ell_0 = 2\pi/V_i$.

Let us use the label $k$ for the material (carrying information not only on the composition through the neutrino potentials but on the layer thickness as well). Then the generic amplitude $A_{\rho\sigma}^{\{k\}}$ that a $\rho$-neutrino enters and a $\sigma$-neutrino leaves layer $\{k\}$ is given for the two neutrino types $\alpha$ and $\beta$ by

$$A_{\alpha\alpha}^{\{k\}} = \cos \varphi_k + i \cos 2\theta_k \sin \varphi_k,$$

$$A_{\beta\beta}^{\{k\}} = \cos \varphi_k - i \cos 2\theta_k \sin \varphi_k, \tag{1.9}$$

$$A_{\alpha\beta}^{\{k\}} = A_{\beta\alpha}^{\{k\}} = -i \sin 2\theta_k \sin \varphi_k.$$

Since we shall mainly be concerned with total conversion it is useful for later comparison to give here the length $X_k$ for maximum conversion in a single layer of material $k$, immediately found from the last of these equations. The maximum possible conversion is realized for a length $X_k$ such that the factor $\sin \varphi_k = 1$, i.e.,

$$X_k = \frac{4E}{\Delta m_k^2} \frac{\pi}{2} = \xi_k \frac{\Delta m^2}{\Delta m_k^2} \frac{\pi}{V_k} = \begin{cases} \xi_k \dfrac{\pi}{V_k} & \text{far below MSW} \\[2ex] \cot 2\theta \dfrac{\pi}{V_k} & \text{at MSW} \\[2ex] \dfrac{\pi}{V_k} & \text{far above MSW} \end{cases} \tag{1.10}$$

The maximum conversion probability is $\sin^2 2\theta_k$, which is $\sin^2 2\theta$ far below MSW, unity at MSW, and $\sin^2 2\theta/\xi_k^2$, asymptotically small, far above MSW.

The smallest value of energy and hence of $\xi_k$ leads to the smallest length $X_k$. At the same time, it is easiest to detect energetic neutrinos. It is therefore helpful to have some idea of the energies that are involved for neutrinos. We note that when we use the number $V_k = 6 \times 10^{-9}$ cm appropriate for earth [2], then

$$\xi_k = (E/(\Delta mc^2)^2) \times 2.5 \times 10^{-13} \text{ eV}$$

For $\Delta mc^2 = \text{O}(10^{-3})$ eV, $\xi_k$ is on the order of 1 for $E = \text{O}(4 \text{ MeV})$. One can think of this energy as roughly the dividing line for whether one is below or above MSW, although of course the precise MSW energy depends not only on $\theta$ but on the correct value of $\Delta m^2$ as well, and the latter number is not yet fully understood [10].

***Propagation through layers***. When there are layers of matter with differing densities, then interference is possible. The history of the effects of passage through repeated layers on mixing is in fact extensive, beginning perhaps with a discussion of neutron-antineutron oscillations in a nonuniform magnetic field [11], then continuing with work on neutrinos by Ermilova et al. [12] and later by Akhmedov [13] in a long series of papers. These latter references describe what the phenomenon of parametric amplification; in fact the work of refs. [7-9] can be seen as a special case of this treatment [14]. However, the approaches taken in refs. [7-9] and in this paper are oriented in a fashion sufficiently different to make them worth independent consideration.

Let us consider two layers oriented perpendicular to a neutrino beam, the first labeled {1} and the second {2}. Then, as pointed out in [7], the amplitude $A_{\alpha\beta}$ for passage through two successive layers {1} and {2} (what we refer to as a bilayer) contains two terms, and these terms can interfere:

$$\begin{aligned} A_{\alpha\beta} &= A_{\alpha\beta}^{\{1\}} A_{\beta\beta}^{\{2\}} + A_{\alpha\alpha}^{\{1\}} A_{\alpha\beta}^{\{2\}} \\ &= \sin\left(2\theta_2 - 2\theta_1\right) \sin\varphi_1 \sin\varphi_2 \\ &\quad -i\left\{\sin 2\theta_1 \sin\varphi_1 \cos\varphi_2 + \sin 2\theta_2 \cos\varphi_1 \sin\varphi_2\right\} \end{aligned} \qquad (1.11)$$

One can see immediately that the structure of this amplitude is not that of the single layer; for example, the third of Eqs. (1.9) is purely imaginary. In particular it is possible for this amplitude to have magnitude one—total conversion—over a wide range of the parameter space.

The work of reference [7] approaches total conversion for a two-channel problem, with a beam initially of type $\alpha$ passing through a double layer, through the probability condition $|A_{\alpha\beta}|^2 = 1$. In a rather involved calculation it is shown using this condition that total conversion occurs for layer thicknesses such that

$$\begin{aligned} y_1^2 &\equiv \tan^2\varphi_1 = \frac{-\cos 2\theta_2}{\cos 2\theta_1 \cos\left(2\theta_2 - 2\theta_1\right)} \\ y_2^2 &\equiv \tan^2\varphi_2 = \frac{-\cos 2\theta_1}{\cos 2\theta_2 \cos\left(2\theta_2 - 2\theta_1\right)} \end{aligned} \qquad (1.12)$$

in regions of $\theta_1$ and $\theta_2$ where the right hand sides are positive. (Note our definition $y_i \equiv \tan\varphi_i$.)

In the two-channel problem Eq. (1.12) is in fact more simply approached through the amplitude condition $A_{\alpha\alpha} = 0$. This amplitude is given by

$$\begin{aligned}
A_{\alpha\alpha} &= A_{\alpha\alpha}^{\{1\}} A_{\alpha\alpha}^{\{2\}} + A_{\alpha\beta}^{\{1\}} A_{\beta\alpha}^{\{2\}} \\
&= \cos\varphi_1 \cos\varphi_2 - \cos\left(2\theta_2 - 2\theta_1\right)\sin\varphi_1 \sin\varphi_2 \\
&\quad - i\left\{\cos 2\theta_1 \sin\varphi_1 \cos\varphi_2 + \cos 2\theta_2 \cos\varphi_1 \sin\varphi_2\right\}.
\end{aligned} \tag{1.13}$$

This amplitude is zero when both its real and imaginary parts vanish, representing two conditions for the two angles $\varphi_1$ and $\varphi_2$ and hence for the layer thicknesses if all other physical parameters are given. One can see immediately from Eq. (1.13) that these conditions are simply written as conditions for $y_1$ and $y_2$, namely

Real part = 0: $\qquad 1 - y_1 y_2 \cos(2\theta_2 - 2\theta_1) = 0.$ $\qquad$ (1.14a)

Imaginary part = 0: $\qquad y_1 \cos 2\theta_1 + y_2 \cos 2\theta_2 = 0.$ $\qquad$ (1.14b)

From these equations, quadratic in the $y_i$, one immediately arrives at the solutions given in Eqs. (1.12).

## II. Passage through multiple layers

The amplitude $A_{\alpha\alpha}$ for the survival of neutrino type $\alpha$ through multiple layers is developed in a straightforward way from the single-layer amplitudes of Eq. (1.9). In the two-family problem, this amplitude is an element of a $2 \times 2$ matrix resulting from the multiplication of two primitive (single layer) $2 \times 2$ matrices. The generalization to more than two layers is straightforward. We give here the cases of three and four layers as examples; in each case we give the conditions for $A_{\alpha\alpha} = 0$.

For three layers, the conditions that the real and imaginary parts of $A_{\alpha\alpha} = 0$ are, respectively,

$$1 - \sum_{\substack{i,j=1 \\ i<j}}^{3} y_i y_j \cos\left(2\theta_i - 2\theta_j\right) = 0 \tag{2.1a}$$

$$\sum_{i=1}^{3} y_i \cos 2\theta_i - y_1 y_2 y_3 \cos\left(2\theta_1 - 2\theta_2 + 2\theta_3\right) = 0. \tag{2.1b}$$

For four layers the respective conditions are

$$1 - \sum_{\substack{i,j=1 \\ i<j}}^{4} y_i y_j \cos\left(2\theta_i - 2\theta_j\right) + y_1 y_2 y_3 y_4 \cos\left(2\theta_1 - 2\theta_2 + 2\theta_3 - 2\theta_4\right) = 0 \tag{2.2a}$$

$$\sum_{i=1}^{4} y_i \cos 2\theta_i - \sum_{\substack{i,j,k=1 \\ i<j<k}}^{4} y_i y_j y_k \cos\left(2\theta_i - 2\theta_j + 2\theta_k\right) = 0. \tag{2.2b}$$

These two examples are sufficient to understand the more general cases. The only important feature to note here is that for more than two layers the two conditions that $A_{\alpha\alpha}$ vanish are insufficient to determine uniquely the $y_i$ and hence the layer thicknesses.

*Repeated layers*. A solvable case is that of alternating layers with every other layer identical to its partners. In other words, we have exactly repeating layer pairs or repeating layer pairs plus a last layer identical to the first. If we label $N$ as the total number of layers, then these possibilities correspond to $N$ even and $N$ odd, respectively. The number of bilayers is $n = [N/2]$, where the square bracket indicates the largest integer in $N/2$. We found no especially interesting solutions for the odd $N$ case and will make only passing comments on it.

For even $N$, we have in mind ultimately a situation in which the first member of a bilayer is vacuum and the second is a given thickness of a dense material, but we treat the more general situation of a separate potential difference for each layer. In this case $\Delta m_1{}^2 = \Delta m_3{}^2 = \Delta m_5{}^2 = \ldots$; $\Delta m_2{}^2 = \Delta m_4{}^2 = \ldots$; $\varphi_1 = \varphi_3 = \ldots$; and so forth, so that we have only the subscripts 1 and 2. The two conditions for real and the imaginary part will now determine $y_1$ and $y_2$.

We give a series of explicit results for the conditions for total conversion for multiple bilayers in the appendix. We remark here that the $2 \times 2$ matrix $A$ that gives the amplitude for the passage through $n$ bilayers can be written as a factor $(\cos\varphi_1 \cos\varphi_2)^n$ times a remaining matrix $A'$. Since the conditions refer to the vanishing only of the $\alpha\alpha$-component of $A$, we derive the (necessary and sufficient) conditions from $A'_{\alpha\alpha} = 0$. These are the conditions given in the appendix.

The calculations presented in the appendix reveal two important feature that we shall assume to be general: First, the imaginary part of the amplitude $A'_{\alpha\alpha}$ contains a single factor of the combination

$$F \equiv y_1\cos2\theta_1 + y_2\cos2\theta_2 \qquad (2.3)$$

This will turn out to be quite useful, as we shall see below. Second, aside from this single factor, the modified mixing angles $\theta_i$ appear in $A'_{\alpha\alpha}$ only in the combination $\delta$ defined by

$$\delta \equiv 2\theta_1 - 2\theta_2. \qquad (2.4)$$

*Conditions for total conversion*. In the case of the single bilayer, the imaginary part in particular vanishes only if the factor $F$ defined by Eq. (2.3) vanishes. For more than one bilayer, either $F$ or its coefficient could vanish. Let us consider the latter possibility for some low order examples.

For $n = 2$, the imaginary part of the amplitude is given by Eq. (A.2b), and we want to consider the possibility that the second term vanishes, i.e. that $1 - y_1y_2\cos\delta = 0$. This gives $y_2 = (y_1\cos\delta)^{-1}$, and when this is substituted into the real part, Eq. (A.2a), we find the condition

$$1 + y_1^2 + \frac{1}{\cos^2\delta} + \frac{1}{y_1^2\cos^2\delta} = 0\,.$$

But each term on the right side of this expression is positive, so we do not have a solution.

For $n = 3$, we consider the possibility that imaginary part vanishes because the expression in curly brackets of Eq. (A.3b) vanishes. But three times the curly bracket in Eq. (A.3b) + the left side of Eq.(A.3a) (the real part for $n = 3$) is

$$-3\left(y_1^2 + 1\right)\frac{y_1^2\cos^2\delta + 1}{y_1^2\cos^2\delta}$$

and, as for its analog in $n = 2$, this quantity cannot vanish.

Although once again we do not have a general proof, it is reasonable that the only way for the imaginary part of the amplitude $A'_{\alpha\alpha}$ to be zero is with the condition that $F$ vanishes. Using this condition we can make an arbitrary $n$ generalization for the form of the amplitude at the total conversion point. To do so we write the amplitude for passage through a single bilayer in canonical form, namely

$$B = \begin{bmatrix} e^{i\omega_{\alpha\alpha}}\cos\eta & e^{i\omega_{\alpha\beta}}\sin\eta \\ e^{i\omega_{\beta\alpha}}\sin\eta & e^{i\omega_{\beta\beta}}\cos\eta \end{bmatrix} \tag{2.5}$$

The parameters of this "unit cell" amplitude are determined by comparison to the explicit result,

$$B = A^{\{1\}}A^{\{2\}}$$

where the single layer amplitudes $A^{\{k\}}$ are given by Eq. (1.9). Using Eq. (1.9), we find that $B$ has the more restrictive form

$$B = \begin{bmatrix} e^{i\omega_{\alpha\alpha}}\cos\eta & e^{i\omega_{\alpha\beta}}\sin\eta \\ -e^{-i\omega_{\alpha\beta}}\sin\eta & e^{-i\omega_{\alpha\alpha}}\cos\eta \end{bmatrix} \tag{2.6}$$

with the three bilayer parameters of this expression, $\eta$, $\omega_{\alpha\alpha}$, and $\omega_{\alpha\beta}$, given in terms of the single layer parameters by

$$\cos\eta = \frac{1}{\sqrt{\left(y_1^2+1\right)\left(y_2^2+1\right)}}\left(1-y_1y_2\cos\delta\right) = \cos\varphi_1\cos\varphi_2\left(1-y_1y_2\cos\delta\right) \tag{2.7}$$

$$\tan\omega_{\alpha\alpha} = \frac{y_1\cos 2\theta_1 + y_2\cos 2\theta_2}{1-y_1y_2\cos\delta} \tag{2.8}$$

$$\tan\omega_{\alpha\beta} = -\frac{y_1\sin 2\theta_1 + y_2\sin 2\theta_2}{y_1y_2\sin\delta}. \tag{2.9}$$

This expression simplifies further if we apply the condition that for $n$ bilayers the factor of Eq. (2.3) is zero at the total conversion point:

$$\omega_{\alpha\alpha} = 0 \quad \text{for total conversion} \tag{2.10}$$

(The denominator of Eq. (2.8) is not independently zero except for the single bilayer case.)

With this condition, the $n$ bilayer amplitude with total conversion of the $\alpha$ beam becomes

$$B^n = \begin{bmatrix} \cos n\eta & e^{i\omega_{\alpha\beta}}\sin n\eta \\ -e^{-i\omega_{\alpha\beta}}\sin n\eta & \cos n\eta \end{bmatrix}. \tag{2.11}$$

In turn, we see immediately that for total conversion

$$\cos(n\eta) = 0. \tag{2.12}$$

In turn Eq. (2.12) gives

$$\eta = \frac{(2m+1)\pi}{2n}, \qquad m = 0,1,\ldots n-1. \tag{2.13}$$

The pair of conditions that $F$ and $\cos(n\eta)$ each vanish provide us with two equations for the angles $\varphi_1$ and $\varphi_2$, i.e., for the layer widths $x_1$ and $x_2$. We shall describe the solution to these equations in the next section.

### III. Total conversion in a repeated multilayer system

We apply the simultaneous conditions $F = 0$ and $\cos n\eta = 0$ to determine $y_i = \tan\varphi_i$ here. The $F = 0$ condition determines $y_1$ in terms of $y_2$. The quantity $y_2$ is determined in terms of the angle $\eta$ by the inversion of Eq. (2.7), which is a quadratic equation for $y_2$ in terms of $\cos\eta$. Since $\cos n\eta$ is an $n$th order polynomial in $\cos\eta$, there are $2n$ solutions for $y_2$. Because $N = 2n$, this matches the number of solutions coming from the $N$th order polynomial for $y_2$ coming from the original real part equations, as described below Eqs. (2.10). Thus we find all the solutions in this way. Because we would like to minimize the thickness of the material layers, we shall be interested in small $y_2$ solutions, and we shall see that this corresponds to small values of $\eta$.

If we define

$$z_i \equiv y_i \cos 2\theta_i, \tag{3.1}$$

then the condition that $F$ vanish reads $z_1 + z_2 = 0$. In turn this means that

$$z_1^2 = z_2^2 \equiv z^2. \tag{3.2}$$

Equation (2.7) now becomes

$$\cos\eta = \frac{1}{\sqrt{\left(z^2 + \cos^2 2\theta_1\right)\left(z^2 + \cos^2 2\theta_2\right)}}\left(\cos 2\theta_1 \cos 2\theta_2 + z^2 \cos\delta\right) \tag{3.3}$$

One can see quickly that for small $\eta$, for which $\cos\eta \to 1$, Eq. (3.3) becomes homogeneous in $z^2$, and so has solutions at $z^2 = 0$. For more detail, we consider separately different regimes of the MSW parameter $\xi_i$. In doing so it is simplest to treat our the layer labeled 1 as a layer of vacuum ($\xi_1 = 0$). It is straightforward to generalize to a bilayer consisting of two different materials each with nonzero values of $\xi_i$.

**$\xi_2$ small (below MSW resonance).** We treat $\xi_2$ as a small perturbation, with $\Delta m_2^2 \cong \Delta m^2$ and $\theta_2 \cong \theta$. We have

$$\cos\delta \cong 1 - (1/2)\xi_2^2 \sin^2 2\theta,$$

and hence Eq. (3.3) becomes to leading order in $\xi_2$

$$f \equiv 1 - \cos\eta \cong \frac{y_2^2\left(1 + y_2^2 \cos^2 2\theta\right)\tan^2 2\theta}{2\left(1 + y_2^2\right)^2}\xi_2^2.$$

With the definition $\alpha \equiv \dfrac{\xi_2^2}{2f}$, this equation reads

$$y_2^2\left(1 + y_2^2 \cos^2 2\theta\right)\alpha \tan^2 2\theta = \left(1 + y_2^2\right)^2. \tag{3.4}$$

The right side of this quadratic equation for $y_2^2$ is O(1) or larger, so there are no solutions unless $\alpha = $ O(1), i.e., $f = $ O($\xi_2^2$). But by comparison with Eq. (2.13), we see that for small values of $m/n$,

$$f \cong \frac{1}{2}\frac{\left(2m+1\right)^2 \pi^2}{4n^2}. \tag{3.5}$$

By choosing $n$ large enough, or more particularly $m/n$ small enough, we can imagine choosing $\alpha = $ O(1). The equation for $y_2^2$ will then have small positive solutions.

The formal solution of Eq. (3.4) is

$$y_{2\pm}^2 = \frac{\pm \sin^2 2\theta \sqrt{\alpha\left(\alpha - 4\cos^2 2\theta\right)} - 2\cos^2 2\theta + \alpha \sin^2 2\theta}{2\cos^2 2\theta \left(1 - \alpha \sin^2 2\theta\right)}. \tag{3.6}$$

We see immediately that there is no (real) solution unless $\alpha > 4\cos^2 2\theta$. The suitable (both small and positive) solutions correspond to the minus branch $y_{2-}^2$, and it is this solution that we look at henceforth. The solutions are smooth as we pass through the point where the denominator factor $1 - \sin^2 2\theta = 0$ and are simple for $\alpha$ greater than or equal to O(1) for the $\theta = 0.7$ case that we use below for illustration. Indeed, in this range we can use the very accurate approximation

$$y_{2-}^2 \cong \frac{\cot^2 2\theta}{\alpha} + \left(2 - \cos^2 2\theta\right)\left(\frac{\cot^2 2\theta}{\alpha}\right)^2 + \ldots$$

all the way to small values of $\alpha$. That is because the expansion is in $\cot^2 2\theta/\alpha$, something that follows from the fact that it is $\alpha \tan^2 2\theta$ that appears in the original equation. Thus we can write our solution in the form

$$y_2 = \frac{\cot 2\theta}{\sqrt{\alpha}} \to \frac{\pi \cot 2\theta}{2n\xi_2},$$

where in the last step we have chosen a large $n$ solution with $m = 0$. Under the assumption that $\cot 2\theta$ is small enough, we can replace $y_2$ by $\varphi_2$, and solve for $x_2$:

$$x_2 = \frac{\pi}{nV_2}\cot 2\theta.$$

The total amount of material $nx_2$ is less than the corresponding amount of a single layer of material 2 [Eq. (1.10)] only in the circumstance that $\cot 2\theta$ is very small.

There is a second problem in this region of $\xi_2$. Once we have found $y_2$, then $y_1$, and hence the thickness of layer 1, is determined through the condition

$$0 = y_1 \cos 2\theta_1 + y_2 \cos 2\theta_2$$

But in this region $\theta_1 \cong \theta_2 \cong \theta$. We would then require $y_1 = -y_2$, or, assuming that $y_2$ is sufficiently small that $\varphi_2 \cong \tan\varphi_2 = y_2$,

$$\varphi_1 \cong \tan^{-1}(-\varphi_2)$$

Since the distances $x_2$ must be positive, the only way we can satisfy this condition is to take $\varphi_1 \cong 2\pi - \varphi_2$, or in other words,

$$x_1 = \frac{8\pi E}{\Delta m^2} - x_2. \tag{3.7}$$

The first term on the right side of this expression, which is the oscillation length in vacuum, is not necessarily small. The multiple bilayer arrangement offers no advantages below MSW.

**$\xi_2$ at MCW resonance**. We have $\theta_1 = \theta$ and $\varphi_1 = (\Delta m^2/4E)x_1$. For medium 2, $\xi_2 = \cos 2\theta$, and $\sin 2\theta_2 = 1$ or $2\theta_2 = \pi/2$ and $\cos 2\theta_2 = 0$. (In fact we shall assume that we are a little above MSW resonance, so that $\cos 2\theta_2$ is small and negative. This helps to clarify limits.) The angle $\varphi_2 = \dfrac{\Delta m_2^2}{4E} x_2 = \dfrac{\Delta m^2 \sin 2\theta}{4E} x_2$. We also have $\delta = \pi/2 - 2\theta$ and $\cos\delta =$ $\sin 2\theta$. In this limit the relation between $y_2$ and $\eta$ of Eq. (3.3) becomes

$$\cos\eta = \frac{1}{\sqrt{y_2^2 + 1}} \tag{3.8}$$

The relevant solution to Eq. (3.8) is

$$y_2 = \tan\eta. \tag{3.9}$$

or, since $y_2 = \tan\varphi_2$,

$$\varphi_2 = \eta = \frac{\pi(2m+1)}{2n}. \tag{3.10}$$

Since $\cos 2\theta_2$ is small and negative, $y_1$ is satisfactorily positive and is also small:

$$y_1 = y_2|\cos 2\theta_2|/\cos 2\theta. \tag{3.11}$$

When we calculate the layer thicknesses we see why this case is of no special interest. Recall [Eq. (1.10)] that at MSW resonance a single layer of thickness $X_2 = (\pi/V_2)$ $\cot 2\theta$ of material 2 gives total conversion. In comparison, Eq. (3.10) shows us that the minimum value of $\varphi_2$ for an $n$-bilayer system occurs for $m = 0$, in which case we have $\varphi_2 = \pi/(2n)$, or

$$x_2 = \frac{\pi}{2n} \frac{4E}{\Delta m_2^2} = \frac{1}{n} \frac{\pi}{V_2} \cot 2\theta. \tag{3.13}$$

Thus the total amount $nx_2$ of material 2 is exactly the amount needed for the single layer. Moreover in the MSW limit, $\cos 2\theta_2$ is zero, so that from Eq. (3.11) $y_1$ and hence the total thickness of vacuum $nx_1$ vanishes. The entire system limits to a single layer of material 2.

We have been able to find no quantity associated with $n$ bilayers that scales to any experimental advantage in the MSW limit.

**$\xi_2$ large (above MSW resonance)**. Again we assume that the medium labeled 1 is vacuum, $\xi_1 = 0$. For $\xi_2 >> \cos\theta$ and $\sin\theta$, $\sin^2 2\theta_2 \cong \sin^2 2\theta/\xi_2^2 \to 0$, with $\sin 2\theta_2$ positive and $\cos 2\theta_2$ approximately $-1$. The fact that $\sin 2\theta_2$ is small means [Eq. (1.9)] that one can at best have very little conversion in a single layer of material 2. Thus the very possibility of total conversion makes this limit interesting.

The relation (3.3) between $y_2$ and $\eta$ gives solutions independent of $\xi_2$ in the large $\xi_2$ limit, namely

$$y_{2\pm} = \pm \frac{\sin\eta\cos 2\theta}{\sqrt{\left|\cos^2 2\theta - \cos^2 \eta\right|}}.$$

The positive $y_2$ solution is then simply

$$y_2 = \frac{|\sin\eta|\cos 2\theta}{\sqrt{\left|\cos^2 2\theta - \cos^2 \eta\right|}} \tag{3.13}$$

Before we deal with the issue of many bilayers, let us consider the case of a single bilayer. We show here that total conversion may not be possible in the single bilayer, although it will always be possible for $n \geq 2$. The original total conversion conditions for the single bilayer are given by Eqs. (1.14). If we take $y_1$ from the second of these equations and substitute into the first, we find an equation for $y_2$, namely

$$1 + y_2^2 \frac{\cos 2\theta_2 \cos\left(2\theta - 2\theta_2\right)}{\cos 2\theta} = 0 \qquad (3.14)$$

Well above MSW, $2\theta_2 = \pi - \varepsilon$, and expanding to leading order in $\varepsilon$ gives

$$y_2^2 = -\frac{1}{1 - \varepsilon \tan 2\theta} \, . \qquad (3.15)$$

For this equation to have a valid (positive) solution, one requires that $\varepsilon \tan 2\theta > 1$, and this will not always hold; indeed it can hold only in a decreasing domain of $\theta$ as $E$ becomes larger ($\varepsilon \to 0$). This situation is illustrated in the numerical example of the next section. It is not difficult to show that there will always be a total conversion solution in this limit for two or more bilayers.

Let us turn next to the case of many bilayers. Given that we are interested in the case of small $\eta$, and supposing that $\cos 2\theta$ is much larger than $\sin \eta$, we can replace the denominator in this expression by $\sin 2\theta$, and our solution becomes

$$y_2 = \cot 2\theta \sin \eta. \qquad (3.16)$$

In turn, Eq. (3.11) gives us the (small) value of $y_1$, namely $y_1 = y_2 |\cos 2\theta_2|/\cos 2\theta \cong \sin \eta / \sin 2\theta$. (In the numerical example treated in the next section, we chose $\theta = 0.7$, in which case $\cot 2\theta \cong 0.17$, while $\sin 2\theta \cong 0.98$.)

We have $\Delta m_2^2 \cong \Delta m^2 \, \xi_2$ in this limit. We also choose the minimum value $\pi/(2n)$ for $\eta$, and expand for small $\eta$. Then we compute from our results for $y_1$ and $y_2$ the total amounts of material 2 and of vacuum space to be, respectively,

$$nx_2 = \frac{\pi}{V_2} \cot 2\theta \quad \text{and} \quad nx_1 = \frac{\pi}{V_2} \xi_2 \csc 2\theta \, . \qquad (3.17)$$

These numbers should be compared to the length of material $X$ needed for *maximum* conversion in the large $\xi_2$ limit, namely [Eq. (1.10)] $X = \pi/V_2$. We see that if the amount of material is the controlling issue one can gain considerably, in that one may have $nx_2 << X$. However, $nx_1 >> X$, so that if the total length of the experiment is the controlling issue this limit is not useful. We should also recall that the maximum conversion in a single layer of width $X$ is asymptotically, so the very possibility of total conversion is an attractive feature of the multiple bilayer arrangement.

## IV. Numerical example

As indicated by the discussion of the previous section, the most interesting cases to look at are those for which $\xi_2$ puts one above the MSW resonance. We present two numerical illustrations here, each for the arbitrary value of $\theta = 0.7$, corresponding to a large degree of mixing. Our bilayer consists of a layer of vacuum followed by a layer of a material 2 for which the potential is given by $V_2 = 6 \times 10^{-9}$ cm [2]. In the first example we assume the energy and masses are such that one is slightly above MSW and in the second example one is far above MSW. Our strategy is to first allow the possibility of total conversion by fixing the thickness $x_1$ of the first layer in terms of the second layer through the condition that $F$, as defined in Eq. (2.3), vanish, i.e. through Eq. (3.11). We then plot the probability for nonconversion as a function of the *total* width $X_2 = nx_2$ of the material layers for various numbers $n$ of bilayers, including the single bilayer.

For the first example, we suppose that we are slightly above MSW resonance, $2\theta_2 = \pi/2 + 0.02$. Figure 1 shows the total length of material versus the probability of nonconversion (i.e., total conversion is a zero in this plot) for $n = 1, 2, 3$, and 6. There is very little dependence on the number of layers. For the parameters used one can directly locate the first large-$n$ zero [Eq. (3.13)], and it matches the numerical value on the plot precisely. We have also made a variation on this calculation, in which we have shifted $x_1$ from the zero-determining value, and we have observed the zero fill in as the shift increases, verifying that the conversion is no longer total.

As a second example we suppose that we are far above MSW, $2\theta_2 = \pi - 0.02$. Figure 2 again shows the probability of nonconversion for $n = 1, 2, 3$, and 6 as a function of the total amount of material used. In this case the single bilayer does not give total conversion. One can see the zero move to the left (less material) as n increases, with the overall pattern quite distinctly dependent on $n$.

## V. Comments

We have concentrated here on the possibility of total conversion of neutrinos in multilayer systems. It would appear that the technique is more interesting at high energies. If these effects are ever to play a role in experiments it will be important to understand several features that we have not looked at, including in particular the implications of a realistic energy spread and, less importantly, the generalization to three families. The three family calculation in principle has a richer variety of possible outcomes for conversion experiments.

The neutral $K$-system presents another well-known case of oscillation. It differs radically from the neutrino system; among other differences materials in the kaon beam produce absorption as well as forward scattering. This system may be interesting to think about from the point of view taken here.

Finally we remark that there is another class of effects that exploits the fact that the order of layers matters in conversion probabilities. We shall discuss this elsewhere.

## Acknowledgements

We particularly want to thank Peter Kaus for helpful suggestions, and Stephen Gasiorowicz for useful conversations. We also want to thank Dominique Schiff and the members of the LPTHE at Université de Paris-Sud for their generous hospitality. This work is supported in part by the U.S. Department of Energy under grant number DE-FG02 -97ER41027.

# References

1. S. P. Mikheyev and A. Yu. Smirnov, Sov. J. Nucl. Phys. **42**, 913 (1985).

2. L. Wolfenstein, Phys. Rev. **D17**, 2369 (1978).

3. Super-Kamiokande Collaboration, Y. Fukuda *et al*., Phys. Lett. B, **433**, 9 (1998); *ibid*., **436**, 33 (1998); Phys. Rev. Lett. **81**, 1562 (1998); *ibid*., **82**, 1810 (1999); *ibid*., **82**, 2430 (1999); E. Kearns, TAUP97, The 5[th] International Workshop on Topics in Astroparticle and Underground Physics, Nucl. Phys. Proc. Suppl. **70**, 315 (1999); A. Habig for the Super-Kamiokande Collaboration, hep-ex/9903047; K. Scholberg for the Super-Kamiokande Collaboration, hep-ex/9905016, to appear in the Proceedings of the 8[th] International Workshop on Neutrino Telescopes (Venice, Italy, 1999); G. L. Fogli, E. Lisi, A. Marrone, and G. Scioscia, hep-ph/9904465, to appear in the Proceedings of *WIN '99*, 17[th] Annual Workshop on Weak Interactions and Neutrinos (Cape Town, South Africa, 1999).

4. Kamiokande Collaboration, K. S. Hirata et al., Phys. Lett. **B280**, 146 (1992); Y. Fukuda *et al*., Phys. Lett. **B335**, 237 (1994); IMB collaboration, R. Becker-Szendy *et al*., Nucl. Phys. B (Proc. Suppl.) **38**, 331 (1995); Soudan-2 collaboration, W. W. M. Allison *et al*., Phys. Lett. **B391**, 491 (1997); Kamiokande Collaboration, S. Hatekeyama *et al*., Phys. Rev. Lett. **81**, 2016 (1998); MACRO Collaboration, M. Ambrosio *et al*., Phys. Lett. B **434**, 451 (1998); CHOOZ Collaboration, M. Apollonio *et al*., Phys. Lett. B420, 397 (1998).

5. J. N. Bahcall and M. H. Pinsonneault, Rev. Mod. Phys. **67**, 781 (1995); J. N. Bahcall, S. Basu, and M. H. Pinsonneault, Phys. Lett. B **433**, 1 (1998); J. N. Bahcall, P. I. Krastev, and A. Yu. Smirnov, Phys. Rev. **D58**, 096016 (1998); B. T. Cleveland *et al*., Nucl. Phys. B (Proc. Suppl.) **38**, 47 (1995); Kamiokande Collaboration, Y. Fukuda *et al*., Phys. Rev. Lett. **77**, 1683 (1996); GALLEX Collaboration, W. Hampel *et al*., Phys. Lett. **B388**, 384 (1996); SAGE Collaboration, J. N. Abdurashitov *et al*., Phys. Rev. Lett. **77**, 4708 (1996); Liquid Scintillator Neutrino Detector (LSND) Collaboration, C. Athanassopoulos *et al*., Phys. Rev. Lett. **75**, 2650 (1996); *ibid*., **77**, 3082 (1996); Phys. Rev. **C58**, 2489 (1998); Phys. Rev. Lett. **81**, 1774 (1998).

6. G. L. Fogli, E. Lisi, A. Marrone, and G. Scioscia, Phys. Rev. **D59**, 033001 (1999). See also G. L. Fogli, E. Lisi, and A. Marrone, Phys. Rev. **D57**, 5893 (1998) and references therein.

7. M.V. Chizhov and S. T. Petcov, Phys. Rev. Lett. **83**, 1096 (1999).

8. M.V. Chizhov and S. T. Petcov, hep-ph/9903424.

9. M.V. Chizhov, hep-ph/9909439.

10. See for example P. M. Fishbane and P. Kaus, J. Phys. G: Nucl. Part. Phys. **26**, 295 (2000) and references therein.

11. G. D. Pusch, Nuovo Cim. **A74**, 149 (1983).

12. V. K. Ermilova, V. A. Tsarev, and V. A. Chechin, Kr. Soob. Fiz. [Short Notices of the Lebedev Institute] **5**, 26 (1986).

13. E. Kh. Akhmedov, preprint IAE-4470/1, 1987; E. Kh. Akhmedov, Yad Fiz. **47**, 475 (1988) [Sov. J. Nucl. Phys. **47**, 301 (1988)]; P. I. Krastev and A. Yu. Smirnov, Phys. Lett. B **226**, 341 (1989); E. Kh. Akhmedov, Nucl. Phys. **B538**, 25 (1999); E. Kh. Akhmedov, A. Dighe, P. Lipari, and A. Yu. Smirnov, Nucl Phys. **B542**, 3 (1999); M. Chizhov, M. Maris, and S. T. Petcov, hep-ph/9810501; E. Kh. Akhmedov, hep-ph/9903302; E. Kh. Akhmedov, Pramana **54**, 47 (2000).

14. E. Kh. Akhmedov and A. Yu. Smirnov, hep-ph/9910433.

## Figure Captions

Figure 1. The total length of material, in units of $10^8$ cm, in a multiple bilayer system consisting of $n$ alternating slices of vacuum and material versus the probability of nonconversion. Total conversion is a zero in this plot. The material has the density of the earth, and the relative width of the layers is determined so that the factor $F$ of Eq. (2.3) is zero, which guarantees the possibility of complete conversion. The energy of the neutrino beam is such that we are slightly above MSW resonance, $2\theta_2 = \pi/2 + 0.02$. Plots are drawn for $n = 1, 2, 3$, and 6.

Figure 2. The total length of material, in units of $10^6$ cm, in a multiple bilayer system consisting of $n$ alternating slices of vacuum and material versus the probability of nonconversion. Total conversion is a zero in this plot. The material has the density of the earth, and the relative width of the layers is determined so that the factor $F$ of Eq. (2.3) is zero, which guarantees the possibility of complete conversion for $n > 1$. The energy of the neutrino beam is such that we are far above MSW resonance, $2\theta_2 = \pi - 0.02$. Plots are drawn for $n = 1, 2, 3$, and 6.

## Appendix

Here we work through a series of cases of total conversion in $n$ bilayers in order to develop insight to the most general case. For $n = 1$, Eqs. (1.14) apply, although it is useful to repeat them here. Through $n = 4$ we find the conditions for total conversion (the "a" and "b" equations refer respectively to the real and imaginary parts)

$n = 1$:

$$1 - y_1 y_2 \cos\delta = 0 \tag{A.1a}$$

$$y_1 \cos 2\theta_1 + y_2 \cos 2\theta_2 = 0. \tag{A.1b}$$

$n = 2$:

$$1 - y_1{}^2 - y_2{}^2 - 4 y_1 y_2 \cos\delta + y_1{}^2 y_2{}^2 \cos(2\delta) = 0 \tag{A.2a}$$

$$2[y_1 \cos 2\theta_1 + y_2 \cos 2\theta_2][1 - y_1 y_2 \cos\delta] = 0 \tag{A.2b}$$

$n = 3$:

$$1 - 3 y_1{}^2 - 3 y_2{}^2 + 3 y_1{}^2 y_2{}^2 - 3 y_1 y_2 \cos\delta\,[3 - y_1{}^2 - y_2{}^2]$$
$$+ 6 y_1{}^2 y_2{}^2 \cos(2\delta) - y_1{}^3 y_2{}^3 \cos(3\delta) = 0 \tag{A.3a}$$

$$[y_1 \cos 2\theta_1 + y_2 \cos 2\theta_2]\{3 - y_1{}^2 - y_2{}^2 + y_1{}^2 y_2{}^2 - 8 y_1 y_2 \cos\delta$$
$$+ 2 y_1{}^2 y_2{}^2 \cos(2\delta)\} = 0 \tag{A.3b}$$

$n = 4$:

$$1 - 6 y_1{}^2 - 6 y_2{}^2 + y_1{}^4 + y_2{}^4 + 16 y_1{}^2 y_2{}^2 - 2 y_1{}^2 y_2{}^2 (y_1{}^2 + y_2{}^2)$$
$$- 8 y_1 y_2 \cos\delta\,[2 - 2 y_1{}^2 - 2 y_2{}^2 + y_1{}^2 y_2{}^2] + 4 y_1{}^2 y_2{}^2 \cos(2\delta)\,[5 - y_1{}^2 - y_2{}^2]$$
$$- 8\, y_1{}^3 y_2{}^3 \cos(3\delta) + y_1{}^4 y_2{}^4 \cos(4\delta) = 0 \tag{A.4a}$$

$$-2[y_1 \cos 2\theta_1 + y_2 \cos 2\theta_2]\{-2 + 2 y_1{}^2 + 2 y_2{}^2 - 4 y_1{}^2 y_2{}^2$$
$$+ y_1 y_2 \cos\delta\,[10 - 2 y_1{}^2 - 2 y_2{}^2 + y_1{}^2 y_2{}^2] - 6 y_1{}^2 y_2{}^2 \cos(2\delta) + y_1{}^3 y_2{}^3 \cos(3\delta)\} = 0. \tag{A.4b}$$

We can also write systematically pieces of terms in $A'_{\alpha\alpha}$ for general $n$. As examples, the terms in the real part that are proportional to $y^N$ (by $y^p$ we mean in general $y_1{}^q y_2{}^{p-q}$, $q$ positive) take the form

$$(y_1 y_2)^n \cos(n\delta), \tag{A.5}$$

while the terms in the real part that are proportional to $y^{N-2}$ are

$$n\left(y_1 y_2\right)^{n-2}\left\{ y_1 y_2 \frac{\sin\left(n\delta\right)}{\sin\delta} + \frac{y_1{}^2 + y_2{}^2}{2} \frac{\sin\left[\left(n-1\right)\delta\right]}{\sin\delta} \right\}. \tag{A.6}$$

The terms proportional to $y^{N-1}$ in the imaginary part are

$$(y_1 y_2)^n [y_1 \cos 2\theta_1 + y_2 \cos 2\theta_2] \tag{A.7}$$

These terms are the *largest* powers of $y$ possible in both the real and imaginary parts. The real part contains even powers only, with the largest power $y^N$; the imaginary part contains odd powers only, with the largest power $y^{N-1}$.

Finally, we can also develop systematically low powers of $y$ in the imaginary and real parts for arbitrary (large) $n$. A few examples are

Real part, constant terms: $\qquad\qquad\qquad 1$ $\qquad\qquad\qquad\qquad$ (A.8)

Real part, $y^2$ terms: $\qquad\quad (n/2)[2ny_1 y_2 \cos\delta + (n-1)(y_1^2 + y_2^2)]$ $\qquad\quad$ (A.9)

Real part, $y^4$ terms:

$$\frac{n(n-1)}{4!}\left\{\begin{array}{l} 2n(n+1)\,y_1^2 y_2^2 \cos 2\delta + 4n(n-2)\,y_1 y_2 \left(y_1^2 + y_2^2\right)\cos\delta \\ + (n-2)\left[(n-3)\left(y_1^2 + y_2^2\right)^2 + 2(n+3)\,y_1^2 y_2^2\right] \end{array}\right\} \qquad \text{(A.10)}$$

Imaginary part, $y^1$ terms:

$$n[y_1 \cos 2\theta_1 + y_2 \cos 2\theta_2] \qquad\qquad \text{(A.11)}$$

Imaginary part, $y^3$ terms:

$$\frac{n(n-1)}{3}\left[y_1 \cos 2\theta_1 + y_2 \cos 2\theta_2\right]\left[2(n+1)\,y_1 y_2 \cos\delta + (n-2)\left(y_1^2 + y_2^2\right)\right] \quad \text{(A.12)}$$

We have worked out such terms all the way through the $y^8$ terms. We have not, however, found a way to generalize every term.

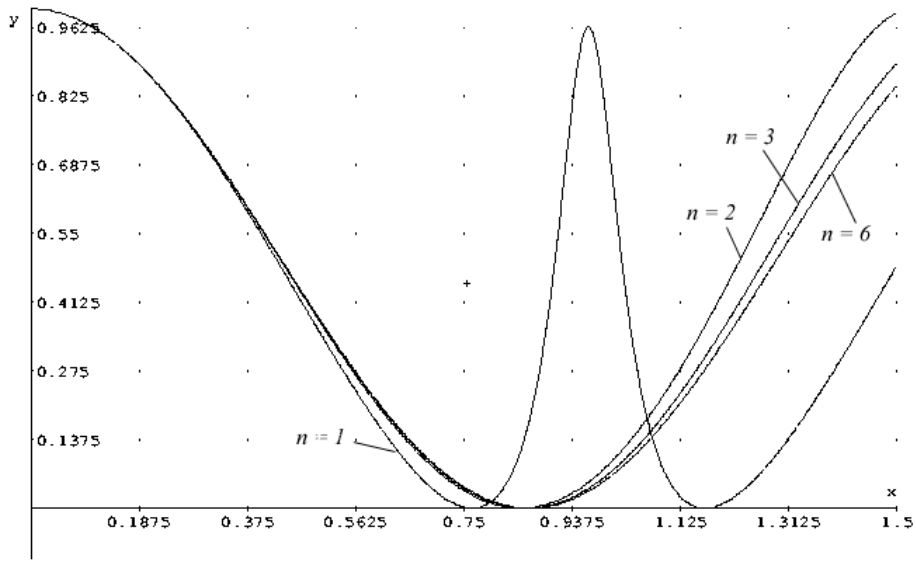

Figure 1

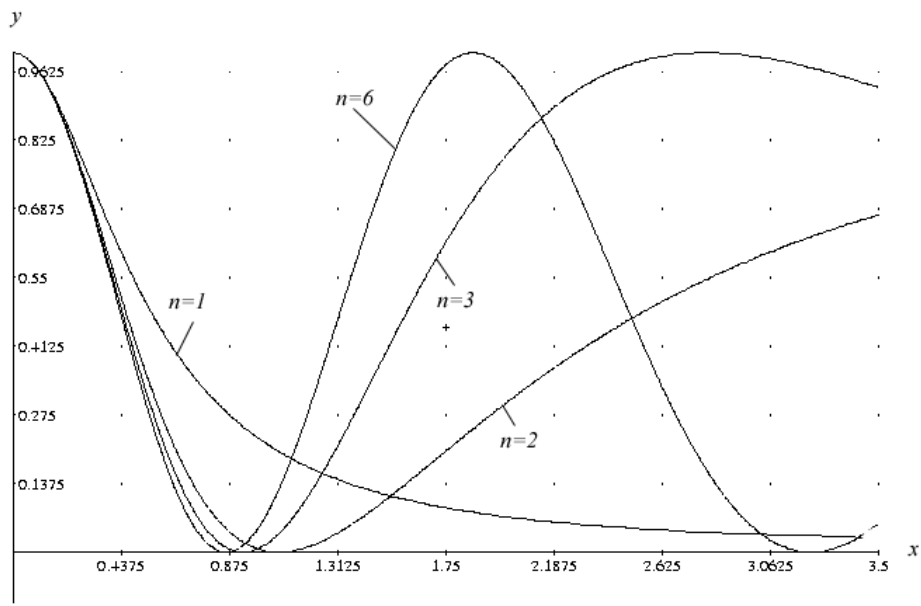

Figure 2